\documentclass[showpacs,twocolumn,preprintnumbers]{revtex4}
\usepackage{amssymb}
\usepackage{amsmath}
\usepackage{graphicx}
\usepackage{dcolumn}
\usepackage{bm}

\begin{document}

\title{Robust Preparation of   GHZ and W States of Three Distant Atoms}
\author{Chang-shui Yu}
\author{X. X. Yi}
\author{He-shan Song}
\email{hssong@dlut.edu.cn}
\author{D. Mei}
\affiliation{Department of Physics,
Dalian University of Technology,\\
Dalian 116024, China}
\date{\today }

\begin{abstract}
Schemes to generate
Greenberger-Horne-Zeilinger(GHZ) and W states of
three distant atoms are proposed in this paper.
The schemes use the effects of quantum statistics
of indistinguishable photons emitted by the atoms
inside optical cavities. The advantages of the
schemes are their robustness against detection
inefficiency and asynchronous emission of the
photons. Moreover, in Lamb-Dicke limit,  the
schemes do not require simultaneous click of the
detectors, this makes the schemes more realizable
in experiments.
\end{abstract}

\maketitle

Entanglement shared by distant parties  can be
employed not only to test quantum nonlocality,
but also is an important physical resource in
quantum information processing (QIP) [1-3].
Although most of the quantum information
protocols concern with   bipartite systems,
multipartite entanglement has also attracted
increasing interest since its potential
applications in QIP. It has been shown [4] that
there exist two inequivalent classes of
multipartite entangled states, i.e.,
Greenberger-Horne-Zeilinger (GHZ) [5] state and W
state [4], which can not be converted to each
other by local operations and classical
communications (LOCC) and   show different
behaviors if one qubit is traced out. Both
classes of entangled state have been shown to
have valuable applications in QIP such as quantum
teleportation [6,7], quantum secret sharing
[8,9], quantum dense coding [10], quantum cloning
machine [11] and so on.

Numerous theoretical proposals [12-16] have been
proposed and many experiments [17-20] have been
conducted  to generate GHZ states and W states.
It has been shown that to entangle distant atoms
(or ions) by using the effect of statistics  of
distinguishable photons is an effective scheme.
For example, in Refs. [21-23], the authors have
presented schemes to entangle two distant qubits
based on the indistinguishability of particles;
Zou et al. [12] have proposed schemes to generate
GHZ and W states of four separate qubits by
different setups; Fidio et al. [13] have also
given an approach to prepare W-type states of
three distant atoms; Duan et al. [14] used the
analogous approach to prepare entangled states of
$N$ atoms. The approach based on
indistinguishability was shown  to have a lot of
advantages, among them the robustness is the most
distinct one. In
this brief report, we propose schemes to prepare GHZ and W states of three \textit{%
distant atoms }based on the indistinguishability of their emitting
photons. The schemes have been shown to be so robust that the
influences of the inefficient detections, the asynchronous emission
of photons have no much effects on the fidelity of the prepared W
and GHZ states.  In particular, in Lamb-Dicke limit, it is not
necessary to require the simultaneous clicks of detectors, which
will relax the condition for  practical realization.

Our schemes work in the same way as the proposal
in Ref. [21], where the authors presented an idea
to entangle two identical $\Lambda -$type
three-level atoms with two degenerate ground
states trapped in two separate cavities. We will
first show that using  different approaches the
similar model can also be used to prepare GHZ and
W states of distant atoms trapped in separate
cavities in terms of simultaneous detections, and
then emphasize that the simultaneous detections
for our schemes are not necessary in Lamb-Dicke
limit. Here we consider three identical $\Lambda
-$type three-level atoms $1$, $2$ and $3$
trapped, respectively, in three spatially
separate optical cavities $A$, $B$ and $C$ which
are all one sided. Every atom has an excited
state $\left\vert e\right\rangle $ and two
degenerate ground states $\left\vert
g_{l}\right\rangle $ and $\left\vert
g_{r}\right\rangle $. The transitions $\left\vert
e\right\rangle \rightarrow \left\vert
g_{l}\right\rangle $ and $\left\vert
e\right\rangle \rightarrow \left\vert
g_{r}\right\rangle $ are strongly coupled to
left- and right-circularly polarizing cavity
modes respectively. The photons leaking out of
every cavity first transmit a quarter wave plate
(QWP), and then pass through two different
setups: One is to prepare GHZ state (See Fig. 1
(a)) and the other is for the preparation of W
state (See Fig. 1 (b)). As shown in Fig. 1 (a),
there are six detectors denoted by $D_{i}^{j}$
with $i=a,b,c$ and $j=F,S$. If any three
detectors with different subscripts
are simultaneously clicked, one will obtain the GHZ states denoted by $%
\left\vert GHZ\right\rangle _{\pm }=\frac{1}{\sqrt{2}}\left( \left\vert
g_{l}\right\rangle _{1}\left\vert g_{l}\right\rangle _{2}\left\vert
g_{l}\right\rangle _{3}\pm \left\vert g_{r}\right\rangle _{1}\left\vert
g_{r}\right\rangle _{2}\left\vert g_{r}\right\rangle _{3}\right) $. If the
three clicked $D_{i}^{j}$s correspond odd number of $F$ (superscript), one
will obtain $\left\vert GHZ\right\rangle _{+}$, otherwise, $\left\vert
GHZ\right\rangle _{-}$. Analogously, in Fig. 1 (b), if any three detectors
are clicked, one will obtain the W state, which consists of two cases: When
both detectors corresponding to BS2 are clicked, one will obtain $\left\vert
W\right\rangle =\frac{1}{\sqrt{3}}\left( \left\vert g_{l}\right\rangle
_{1}\left\vert g_{l}\right\rangle _{2}\left\vert g_{r}\right\rangle
_{3}+\left\vert g_{l}\right\rangle _{1}\left\vert g_{r}\right\rangle
_{2}\left\vert g_{l}\right\rangle _{3}+\left\vert g_{r}\right\rangle
_{1}\left\vert g_{l}\right\rangle _{2}\left\vert g_{l}\right\rangle
_{3}\right) $, otherwise, $\left\vert \tilde{W}\right\rangle =\frac{1}{\sqrt{%
3}}(\left\vert g_{l}\right\rangle _{1}\left\vert g_{r}\right\rangle
_{2}\left\vert g_{r}\right\rangle _{3}+\left\vert g_{r}\right\rangle
_{1}\left\vert g_{r}\right\rangle _{2}\left\vert g_{l}\right\rangle
_{3}+\left\vert g_{r}\right\rangle _{1}\left\vert g_{l}\right\rangle
_{2}\left\vert g_{r}\right\rangle _{3})$ will be obtained.

In the following we will illustrate our approach explicitly. The interaction
Hamiltonian governing the interaction between the trapped $\Lambda -$type
atoms and cavities is given by $H_{I}=\hbar \sum_{k=l,r}\lambda
_{k}(a_{k}\left\vert e\right\rangle \left\langle g_{k}\right\vert
+a_{k}^{\dagger }\left\vert g_{k}\right\rangle \left\langle e\right\vert ),$%
where $l,r$ denoted the left- and right-circularly polarizing cavity modes, $%
a_{k}^{\dagger }$ and $a_{k}$ are the creation and annihilation operators of
photons in the $k$ mode, and $\lambda _{k},$ supposed to be real, is the
coupling constant between the atom and the $k$ mode. If the atom and the
cavity are initially prepared in the excited state $\left\vert
e\right\rangle $ and the vacuum state $\left\vert 0_{l}\right\rangle
\left\vert 0_{r}\right\rangle $, respectively, after the interaction time $t$%
, the total system of atom and cavity will evolve to the state
\begin{equation}
\left\vert \Psi (t)\right\rangle =\cos \Omega t\left\vert e\right\rangle
\left\vert 0_{l}\right\rangle \left\vert 0_{r}\right\rangle -i\sin \Omega
t\left\vert \phi (t)\right\rangle ,
\end{equation}%
with $\left\vert \phi (t)\right\rangle =\frac{1}{\Omega }\left( \lambda
_{l}\left\vert g_{l}\right\rangle \left\vert 1_{l}\right\rangle \left\vert
0_{r}\right\rangle +\lambda _{r}\left\vert g_{r}\right\rangle \left\vert
0_{l}\right\rangle \left\vert 1_{r}\right\rangle \right) $ and $\Omega =%
\sqrt{\lambda _{l}^{2}+\lambda _{r}^{2}}$ supposed to be a given constant.
When photons are passing through the QWP, circularly polarizing photons
become linearly polarizing. Analogous to Ref. [21], we suppose the left- and
right-circularly polarizing photons correspondingly become vertically
(denoted by $V$) and horizontally ($H$) polarizing, i.e. $\left\vert
1_{l}\right\rangle \left\vert 0_{r}\right\rangle \rightarrow \left\vert
V\right\rangle $ and $\left\vert 0_{l}\right\rangle \left\vert
1_{r}\right\rangle \rightarrow \left\vert H\right\rangle $. Furthermore,
because the vacuum state has no contribution to the click of the
photodetectors, the term $\left\vert e\right\rangle \left\vert
0_{l}\right\rangle \left\vert 0_{r}\right\rangle $ in eq. (1) can be safely
neglected for simplification. Therefore, when photon passing through the
QWP, the total state of photon and atom can be written by
\begin{equation}
\left\vert \psi (t)\right\rangle =\frac{1}{\Omega }\left( \lambda
_{l}\left\vert g_{l}\right\rangle \left\vert V\right\rangle +\lambda
_{r}\left\vert g_{r}\right\rangle \left\vert H\right\rangle \right) ,
\end{equation}%
associated with a probability $P_{1}=\sin ^{2}\Omega t$. Later it implies
that photons have passed through QWP if we say photons leak out of cavities.

\begin{figure}[tbp]
\includegraphics[width=9.5cm]{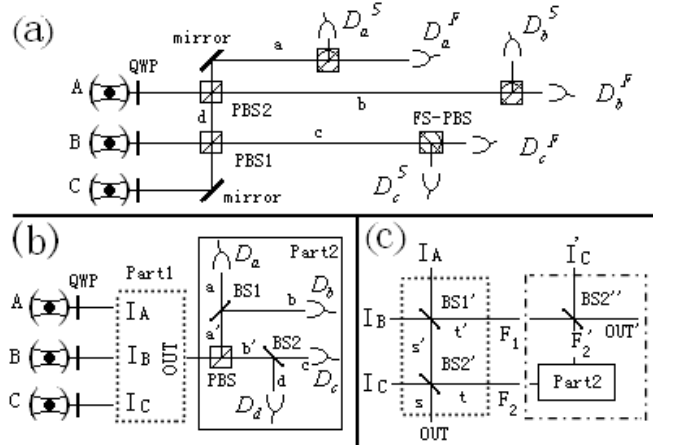}
\caption{(a) Experimental setup for GHZ states. After transmitting quarter
wave plates (QWP), photons leaking out of cavities pass through the
polarizing beam splitters (PBS) and rotated PBS (FS-PBS) and then are
detected by photodetectors. (b) Experimental setup for W states. The setup
includes Part1, bunching system and Part2, detection system. Entering port "I%
$_i$" $i=A,B,C$ and bunched out of port "OUT", photons pass through
PBS and 50/50 beam splitters (BS) and then are detected by
photodetectors. (c) The expanded setups for (b). Dotted box of (c)
sketches an alternate scheme for Part1, where photons passing
through two 50/50 BS are bunched via port "OUT". "F$_1 $" and
"F$_2$" are two expanded ports to improve the success probability,
which can be connected with the dash-dotted box to double the
efficiency of (b). In the dash-dotted box auxillary atom 3' and
cavity C' are introduced via the port "I'$_{C}$". Thus BS2" and BS2'
become two symmetric arms of BS1'.}
\end{figure}

\textit{GHZ states- }See Fig. 1 (a).\textit{\ }Photons leaking out of cavity
$B$ and $C$ will first meet PBS1 which always transmits $H$-polarizing
photons and reflects $V$-polarizing photons (all PBS in the paper work in
this way), hence the input modes $\left\vert V\right\rangle _{B},\left\vert
H\right\rangle _{B}$ will correspondingly change to the output modes $%
\left\vert V\right\rangle _{d},\left\vert H\right\rangle _{c}$, and $%
\left\vert V\right\rangle _{C},\left\vert H\right\rangle _{C}\rightarrow
\left\vert V\right\rangle _{c},\left\vert H\right\rangle _{d}$. Thus the
joint state including cavities $B$, $C$ and atoms $1$, $2$ will become
\begin{eqnarray}
&&\frac{1}{2}\left( \left\vert g_{l}\right\rangle _{2}\left\vert
V\right\rangle _{B}+\left\vert g_{r}\right\rangle _{2}\left\vert
H\right\rangle _{B}\right) \otimes \left( \left\vert g_{l}\right\rangle
_{3}\left\vert V\right\rangle _{C}+\left\vert g_{r}\right\rangle
_{3}\left\vert H\right\rangle _{C}\right)   \notag \\
&\rightarrow &\frac{1}{2}\left( \left\vert g_{l}\right\rangle _{2}\left\vert
g_{l}\right\rangle _{3}\left\vert V\right\rangle _{d}\left\vert
V\right\rangle _{c}+\left\vert g_{l}\right\rangle _{2}\left\vert
g_{r}\right\rangle _{3}\left\vert V\right\rangle _{d}\left\vert
H\right\rangle _{d}\right.   \notag \\
&&\left. +\left\vert g_{r}\right\rangle _{2}\left\vert g_{l}\right\rangle
_{3}\left\vert H\right\rangle _{c}\left\vert V\right\rangle _{c}+\left\vert
g_{r}\right\rangle _{2}\left\vert g_{r}\right\rangle _{3}\left\vert
H\right\rangle _{c}\left\vert H\right\rangle _{d}\right)   \notag \\
&\rightarrow &\frac{1}{\sqrt{2}}\left( \left\vert g_{l}\right\rangle
_{2}\left\vert g_{l}\right\rangle _{3}\left\vert V\right\rangle
_{d}\left\vert V\right\rangle _{c}+\left\vert g_{r}\right\rangle
_{2}\left\vert g_{r}\right\rangle _{3}\left\vert H\right\rangle
_{c}\left\vert H\right\rangle _{d}\right) ,
\end{eqnarray}%
where we suppose $\lambda _{l}=\lambda _{r}$ for GHZ state. As mentioned
above, GHZ state requires that both modes $c$ and $d$ are not idle,
otherwise one of the required detectors can not be clicked, hence we has
discarded the bunching outcome in eq. (3) and preserved the antibunching
outcome with the probability $P_{2}=50\%$ (We discard the bunching outcome
just for simplification, one can not do so which will lead to the same
result). Then the photon of mode $d$ and that leaking out of cavity $A$ will
meet PBS2, which can lead to the following transformation between input
modes and output modes: $\left\vert V\right\rangle _{d},\left\vert
H\right\rangle _{d}\rightarrow \left\vert V\right\rangle _{b},\left\vert
H\right\rangle _{a}$ and $\left\vert V\right\rangle _{A},\left\vert
H\right\rangle _{A}\rightarrow \left\vert V\right\rangle _{a},\left\vert
H\right\rangle _{b}$. As a result, the joint state of the whole system
(three atoms and three cavity modes) follows that%
\begin{eqnarray}
&&\frac{1}{2}\left( \left\vert g_{l}\right\rangle _{1}\left\vert
V\right\rangle _{A}+\left\vert g_{r}\right\rangle _{1}\left\vert
H\right\rangle _{A}\right)   \notag \\
&&\otimes \left( \left\vert g_{l}\right\rangle _{2}\left\vert
g_{l}\right\rangle _{3}\left\vert V\right\rangle _{d}\left\vert
V\right\rangle _{c}+\left\vert g_{r}\right\rangle _{2}\left\vert
g_{r}\right\rangle _{3}\left\vert H\right\rangle _{c}\left\vert
H\right\rangle _{d}\right)   \notag \\
&\rightarrow &\frac{1}{\sqrt{2}}(\left\vert g_{l}\right\rangle
_{1}\left\vert g_{l}\right\rangle _{2}\left\vert g_{l}\right\rangle
_{3}\left\vert V\right\rangle _{a}\left\vert V\right\rangle _{b}\left\vert
V\right\rangle _{c}  \notag \\
&&+\left\vert g_{r}\right\rangle _{1}\left\vert g_{r}\right\rangle
_{2}\left\vert g_{r}\right\rangle _{3}\left\vert H\right\rangle
_{a}\left\vert H\right\rangle _{b}\left\vert H\right\rangle _{c}),
\end{eqnarray}%
where we only preserve the antibunching outcome with the probability $%
P_{3}=50\%$ for the same reason. At last, the three photons with different
modes will, respectively, meet three rotated polarizing beam splitters
(FS-PBS), which change $\left\vert H\right\rangle _{i}$ and $\left\vert
V\right\rangle _{i},i=a,b,c$ into a new frame as $\left\vert H\right\rangle
_{i}=\frac{1}{\sqrt{2}}\left( \left\vert F\right\rangle _{i}-\left\vert
S\right\rangle _{i}\right) $ and $\left\vert V\right\rangle _{i}=\frac{1}{%
\sqrt{2}}\left( \left\vert F\right\rangle _{i}+\left\vert S\right\rangle
_{i}\right) $ and always reflect $S$-polarizing photons and transmit $F$%
-polarizing photons. In the new frame, the state given in eq. (4) can be
written by
\begin{equation}
\sum_{X,Y,Z=F,S}\frac{\left( \left\vert g_{l}\right\rangle _{1}\left\vert
g_{l}\right\rangle _{2}\left\vert g_{l}\right\rangle _{3}\pm \left\vert
g_{r}\right\rangle _{1}\left\vert g_{r}\right\rangle _{2}\left\vert
g_{r}\right\rangle _{3}\right) \left\vert X\right\rangle _{a}\left\vert
Y\right\rangle _{b}\left\vert Z\right\rangle _{c}}{\sqrt{2}},
\end{equation}%
where odd number of $F$ among $X,Y,Z$ corresponds to $"+"$, otherwise, $"-"$%
. Therefore, if three detectors of different mode of $a$, $b$ and $c$ are
clicked simultaneously, the distant atoms $1,2$ and $3$ will collapse to one
of the GHZ states $\left\vert GHZ\right\rangle _{\pm }=$ $\frac{1}{\sqrt{2}}%
\left( \left\vert g_{l}\right\rangle _{1}\left\vert g_{l}\right\rangle
_{2}\left\vert g_{l}\right\rangle _{3}\pm \left\vert g_{r}\right\rangle
_{1}\left\vert g_{r}\right\rangle _{2}\left\vert g_{r}\right\rangle
_{3}\right) $. The maximal probability of getting the state is given by $%
P_{GHZ}=\left( P_{1}\right) ^{3}P_{2}P_{3}$ $=\sin ^{6}\Omega t\times
50\%\times 50\%=25\%$ with $\sin ^{6}\Omega t=1$.

\textit{W state}-Let us turn to Fig. 1 (b). At first, we suppose three
photons leaking out of the three cavities can be directly bunched. An
alternate scheme will be presented later to bunch the three photons. As a
result, when the three photons are transmitted out of the output port "OUT",
the joint state of the whole system is
\begin{equation}
\overset{3}{\underset{i=1}{\otimes }}\frac{1}{\Omega }\left( \lambda
_{l}\left\vert g_{l}\right\rangle _{i}\left\vert V\right\rangle +\lambda
_{r}\left\vert g_{r}\right\rangle _{i}\left\vert H\right\rangle \right)
\end{equation}%
where we have discard the subscripts of cavity modes, because one can not
determine which cavity a photon comes from due to the indistinguishability
of photons. Then they will meet PBS1 which will transform the input modes to
the output modes as $\left\vert V\right\rangle \rightarrow \left\vert
V\right\rangle _{a^{\prime }}$ and $\left\vert H\right\rangle \rightarrow
\left\vert H\right\rangle _{b^{\prime }}$. Thus the state given by eq. (6)
becomes%
\begin{eqnarray}
&&\frac{1}{\Omega ^{3}}\left[ \lambda _{l}^{3}\left\vert g_{l}\right\rangle
_{1}\left\vert g_{l}\right\rangle _{2}\left\vert g_{l}\right\rangle
_{3}\left\vert 3\right\rangle _{a^{\prime }}^{V}+\lambda _{r}^{3}\left\vert
g_{r}\right\rangle _{1}\left\vert g_{r}\right\rangle _{2}\left\vert
g_{r}\right\rangle _{3}\left\vert 3\right\rangle _{b^{\prime }}^{H}\right.
\notag \\
&&+\lambda _{l}^{2}\lambda _{r}\left( \left\vert g_{l}\right\rangle
_{1}\left\vert g_{l}\right\rangle _{2}\left\vert g_{r}\right\rangle
_{3}+\left\vert g_{l}\right\rangle _{1}\left\vert g_{r}\right\rangle
_{2}\left\vert g_{l}\right\rangle _{3}\right.   \notag \\
&&+\left. \left\vert g_{r}\right\rangle _{1}\left\vert g_{l}\right\rangle
_{2}\left\vert g_{l}\right\rangle _{3}\right) \left\vert 2\right\rangle
_{a^{\prime }}^{V}\left\vert 1\right\rangle _{b^{\prime }}^{H}+\lambda
_{l}\lambda _{r}^{2}\left( \left\vert g_{l}\right\rangle _{1}\left\vert
g_{r}\right\rangle _{2}\left\vert g_{r}\right\rangle _{3}\right.   \notag \\
&&+\left. \left\vert g_{r}\right\rangle _{1}\left\vert g_{l}\right\rangle
_{2}\left\vert g_{r}\right\rangle _{3}+\left\vert g_{r}\right\rangle
_{1}\left\vert g_{r}\right\rangle _{2}\left\vert g_{l}\right\rangle
_{3}\right) \left\vert 2\right\rangle _{b^{\prime }}^{H}\left\vert
1\right\rangle _{a^{\prime }}^{V},
\end{eqnarray}%
where $\left\vert 2\right\rangle _{a^{\prime }}^{V}\left\vert 1\right\rangle
_{b^{\prime }}^{H}=\left\vert V\right\rangle _{a^{\prime }}\left\vert
V\right\rangle _{a^{\prime }}\left\vert H\right\rangle _{b^{\prime }}$ with
superscripts denoting polarizing modes and subscripts denoting output port
modes. Finally, the photons will pass through two 50/50 beam splitters BS1
and BS2 which will transform $\left\vert 1\right\rangle _{a^{\prime }}^{V}$
and $\left\vert 1\right\rangle _{b^{\prime }}^{H}$ to the final detecting
modes as $\left\vert 1\right\rangle _{a^{\prime }}^{V}\rightarrow \frac{1}{%
\sqrt{2}}\left( \left\vert 1\right\rangle _{a}^{V}+i\left\vert
1\right\rangle _{b}^{V}\right) $ and $\left\vert 1\right\rangle _{b^{\prime
}}^{H}\rightarrow \frac{1}{\sqrt{2}}\left( \left\vert 1\right\rangle
_{c}^{H}-i\left\vert 1\right\rangle _{d}^{H}\right) $. Hence we have the
following transformations:

\begin{eqnarray}
\left\vert 2\right\rangle _{a^{\prime }}^{V}\left\vert 1\right\rangle
_{b^{\prime }}^{H} &\rightarrow &\frac{1}{2\sqrt{2}}\left( \left\vert
2\right\rangle _{a}^{V}\left\vert 1\right\rangle _{c}^{H}-\left\vert
2\right\rangle _{b}^{V}\left\vert 1\right\rangle _{c}^{H}\right.
+2i\left\vert 1\right\rangle _{a}^{V}\left\vert 1\right\rangle
_{b}^{V}\left\vert 1\right\rangle _{c}^{H}  \notag \\
&-&i\left. \left\vert 2\right\rangle _{a}^{V}\left\vert 1\right\rangle
_{d}^{H}+i\left\vert 2\right\rangle _{b}^{V}\left\vert 1\right\rangle
_{d}^{H}+2\left\vert 1\right\rangle _{a}^{V}\left\vert 1\right\rangle
_{b}^{V}\left\vert 1\right\rangle _{d}^{H}\right)
\end{eqnarray}%
and
\begin{eqnarray}
\left\vert 2\right\rangle _{b^{\prime }}^{H}\left\vert 1\right\rangle
_{a^{\prime }}^{V} &\rightarrow &\frac{1}{2\sqrt{2}}\left( \left\vert
2\right\rangle _{c}^{H}\left\vert 1\right\rangle _{a}^{V}-\left\vert
2\right\rangle _{d}^{H}\left\vert 1\right\rangle _{a}^{V}-2i\left\vert
1\right\rangle _{c}^{H}\left\vert 1\right\rangle _{d}^{H}\left\vert
1\right\rangle _{a}^{V}\right.  \notag \\
&+&i\left. \left\vert 2\right\rangle _{c}^{H}\left\vert 1\right\rangle
_{b}^{V}-i\left\vert 2\right\rangle _{d}^{H}\left\vert 1\right\rangle
_{b}^{V}+2\left\vert 1\right\rangle _{c}^{H}\left\vert 1\right\rangle
_{d}^{H}\left\vert 1\right\rangle _{b}^{V}\right) .
\end{eqnarray}%
Substitute the two transformations into eq. (7), eq. (7) can be written in
terms of the detecting modes of photons. If no two photons are allowed to
click the same detector, the terms where different detectors are clicked can
be given by
\begin{eqnarray}
&&\frac{\lambda _{l}}{\sqrt{2}\Omega }\left\vert W\right\rangle \left(
i\left\vert 1\right\rangle _{a}^{V}\left\vert 1\right\rangle
_{b}^{V}\left\vert 1\right\rangle _{c}^{H}+\left\vert 1\right\rangle
_{a}^{V}\left\vert 1\right\rangle _{b}^{V}\left\vert 1\right\rangle
_{d}^{H}\right)  \notag \\
&&+\frac{\lambda _{r}}{\sqrt{2}\Omega }\left\vert \tilde{W}\right\rangle
\left( \left\vert 1\right\rangle _{c}^{H}\left\vert 1\right\rangle
_{d}^{H}\left\vert 1\right\rangle _{b}^{V}-i\left\vert 1\right\rangle
_{c}^{H}\left\vert 1\right\rangle _{d}^{H}\left\vert 1\right\rangle
_{a}^{V}\right) .
\end{eqnarray}%
The probability of getting such a state from eq. (7) can be given by $%
P^{\prime }=\left\vert \frac{\sqrt{3}\lambda _{l}\lambda _{r}\Omega }{\Omega
^{3}}\right\vert ^{2}\times \frac{2}{3}=\frac{2\lambda _{l}^{2}\lambda
_{r}^{2}}{\Omega ^{4}}$, where one should first normalize eq. (8) and eq.
(9). From eq. (10) a $W-$class state can always be obtained if three
detectors have been simultaneously clicked. If the three clicked detectors
include $D_{a}$ and $D_{b}$, one will obtain $\left\vert W\right\rangle $,
otherwise $\left\vert \tilde{W}\right\rangle $. The maximal probability of
getting $W$ states is $\tilde{P}_{W}=\left( P_{1}\right) ^{3}P^{\prime
}=2\sin ^{6}\Omega t\frac{\lambda _{l}^{2}\lambda _{r}^{2}}{\Omega ^{4}}=%
\frac{1}{2}$ in terms of $\sin ^{6}\Omega t=1$ and $\lambda _{l}=\lambda
_{r} $.

Now we present an alternate scheme sketched as Fig. 1 (c) to bunch the three
photons. Following Fig. 1 (c), one can find that when the two photons
leaking out of cavities $A$ and $B$ pass through BS1', the joint state $%
\frac{1}{\Omega ^{2}}\left( \lambda _{l}\left\vert g_{l}\right\rangle
_{1}\left\vert V\right\rangle _{A}+\lambda _{r}\left\vert g_{r}\right\rangle
_{1}\left\vert H\right\rangle _{A}\right) \otimes \left( \lambda
_{l}\left\vert g_{l}\right\rangle _{2}\left\vert V\right\rangle _{B}+\lambda
_{r}\left\vert g_{r}\right\rangle _{2}\left\vert H\right\rangle _{B}\right) $
will be transformed to the following non-normalized state
\begin{eqnarray}
&&\frac{1}{2}\left( \left\vert \Psi \right\rangle _{t^{\prime }}+\left\vert
\Psi \right\rangle _{s^{\prime }}\right) +i\frac{\lambda _{l}\lambda _{r}}{%
2\Omega ^{2}}\left( \left\vert g_{l}\right\rangle _{1}\left\vert
g_{r}\right\rangle _{2}-\left\vert g_{r}\right\rangle _{1}\left\vert
g_{l}\right\rangle _{2}\right)   \notag \\
&&\times \left( \left\vert 1\right\rangle _{s^{\prime }}^{V}\left\vert
1\right\rangle _{t^{\prime }}^{H}-\left\vert 1\right\rangle _{t^{\prime
}}^{V}\left\vert 1\right\rangle _{s^{\prime }}^{H}\right) ,
\end{eqnarray}%
with $\left\vert \Psi \right\rangle _{i}=\frac{1}{\Omega ^{2}}(\lambda
_{l}^{2}\left\vert g_{l}\right\rangle _{1}\left\vert g_{l}\right\rangle
_{2}\left\vert 2\right\rangle _{i}^{V}+\lambda _{r}^{2}\left\vert
g_{r}\right\rangle _{1}\left\vert g_{r}\right\rangle _{2}\left\vert
2\right\rangle _{i}^{H}+\lambda _{l}\lambda _{r}\left\vert
g_{l}\right\rangle _{1}\left\vert g_{r}\right\rangle _{2}\left\vert
2\right\rangle _{i}^{V,H}+\lambda _{l}\lambda _{r}\left\vert
g_{r}\right\rangle _{1}\left\vert g_{l}\right\rangle _{2}\left\vert
2\right\rangle _{i}^{V,H})$, $i=s^{\prime },t^{\prime }$ being the initial
joint state before inputting BS1. It is obvious that the state can collapse
to $\left\vert \Psi \right\rangle _{i}$ bunching in the output port $%
t^{\prime }$ with the probability $P_{t^{\prime }}=\frac{\Omega ^{4}}{%
2\Omega ^{4}+4\lambda _{l}^{2}\lambda _{r}^{2}}$. By the same algebra, one
can find that the joint state $\left\vert \Psi \right\rangle _{i}\otimes
\frac{1}{\Omega }\left( \lambda _{l}\left\vert g_{l}\right\rangle
_{3}\left\vert V\right\rangle +\lambda _{r}\left\vert g_{r}\right\rangle
_{3}\left\vert H\right\rangle \right) $ can collapse to itself (given in eq.
(6)) bunching in the final port "OUT" with the probability $P_{s}=\frac{%
\Omega ^{4}}{2\Omega ^{4}+8\lambda _{l}^{2}\lambda _{r}^{2}}$. In this case
the maximal probability of getting $W$ states is $P_{W1}=P^{\prime
}P_{t^{\prime }}P_{s}=\frac{\Omega ^{4}}{2\Omega ^{4}+4\lambda
_{l}^{2}\lambda _{r}^{2}}\cdot \frac{\Omega ^{4}}{4\Omega ^{4}+8\lambda
_{l}^{2}\lambda _{r}^{2}}\cdot \frac{2\lambda _{l}^{2}\lambda _{r}^{2}}{%
\Omega ^{4}}=\frac{1}{36}$ with $\lambda _{l}=\lambda _{r}$, $P_{t^{\prime
}}=\frac{1}{3},P_{s}=\frac{1}{6}$. In fact, one can also find that the joint
state of the whole system can be bunched via the port "F2" with the same
probability to $P_{s}$. Hence, if we connect "F2" with the same setup
depicted as "Part 2" in Fig. 1 (b), the probability $P_{W1}$ is doubled.
Analogously, one can find that the initial joint state before BS1 can be
bunched via the port "F1" with probability $P_{t^{\prime }}$. If we
introduce an auxillary atom $3^{\prime }$ trapped in cavity $C^{\prime }$
connected with the input port I$_{C}^{\prime }$, one can finally obtain the $%
W$ states of atoms $A$, $B$ and $C^{\prime }$ (denoted by $W^{\prime }$) by
the setup depicted in the dot-dashed box of Fig. 1 (c). Both setups with and
without "$prime$" are completely identical. Hence, from the viewpoint of the
yield of $W$ states, the doubled probability $P_{W1}$ should be doubled
again. Thus the total probability of getting $W$ states is $\frac{1}{36}%
\times 4=\frac{1}{9}$. Compared with the preparation of the W states [16] of
photons, our probability is higher (than $\frac{3}{32}$).

Both schemes emphasize the simultaneous clicks of detectors. The key is in
that the spirit of our schemes is the indistinguishability. Because
absorption or emission of photons will lead to a recoil of the atom, which
will signal the atom [22], the indistinguishability is destroyed and lead to
no entanglement. However, if our trapped atoms are restricted to operating
in the Lamb--Dicke limit, where the recoil energy does not suffice to change
the atomic motional state [22], the indistinguishability can be preserved
and it is not necessary to require the simultaneous clicks. This in fact
dramatically simplifies the practical operations. To test the validity, one
can first suppose the clicked order of the detectors (such as $%
D_{a}^{j}\rightarrow D_{b}^{j}\rightarrow D_{c}^{j}$ and so on which must
have different subscripts), and then follow the analogous procedures given
above. In this way, so long as the clicked detectors are the same to those
shown by simultaneous detections, one can obtain the same state (We have
tested all the cases).

Here let us briefly discuss the robustness of our schemes. The scheme for
GHZ state needs $\lambda _{l}=\lambda _{r}$, otherwise the final state $%
\left\vert \Phi \right\rangle $ will deviate from $\left\vert
GHZ\right\rangle $. In fact, if $\lambda _{l}/\lambda _{r}=1.1$, the
fidelity is $\left\vert \left\langle \Phi \right. \left\vert
GHZ\right\rangle \right\vert ^{2}>0.98$, which shows slight influence. But
from above derivation for W states, one can find that different $\lambda
_{l} $ and $\lambda _{r}$ only reduce the efficiency, while the fidelity is
not influenced at all. One can also find that the 50/50 beam splitters are
employed in the scheme for $W$ states. If the beam splitters reflect and
transmit photons without equal amplitudes, it is surprising that only the
final efficiency is reduced, but the fidelity can not be changed. What is
more, there exist the following four negative effects: (a) Not all the atoms
are initially prepared in their excited states; (b) Some photons emitting to
the free modes can not be detected; (c) Dark counts of the detectors, or
some photons are absorbed by cavity walls, even optical elements; (d)
Photons are not leaked out simultaneously. However, (a-c) can not lead to
\textit{three} clicked detectors, and (d) will not make the desired three
detectors clicked simultaneously. Therefore, all the cases can be
effectively ruled out with the fidelity of the desired states invariant but
sacrificing the preparation efficiency. However, in Lamb-Dicke limits
besides that (a-c) can be ruled out analogously, (d) gives a positive
contribution to the efficiency. Considering the cavity decay and atomic
spontaneous emission which reduce the efficiency, the same discussions to
those in Ref. [21] are valid, which is not repeated here.

In experimental scenario, our atomic level structure can be achieved by
Zeeman sublevels [24] and has been realized to entangled two atoms [25].
What we used also consists of linear optical elements, and photon detectors,
which has been widely used to entangle photons. In particular, the similar
optical setups has been used to successfully prepare W states of photons in
experiment [17]. Therefore, our schemes are feasible by current technologies.

In conclusion, we have proposed two  schemes to
prepare GHZ states and W states of distant atoms,
respectively, based on the indistinguishability
of photons emitted by the  atoms. Our schemes are
robust against the detection inefficiency, the
asynchronous emission of photons. In particular,
in Lamb-Dicke limit it is not necessary to
require simultaneous click of the detectors,
which will relax the requirement on the physical
realization. The schemes are feasible by current
technologies. They can readily be generalized to
entangle multipartite distant atoms in principle,
but the efficiency would be reduced.

This work was supported by the  National Natural Science Foundation
of China, under Grant No. 10575017 and No. 60578014.

\end{document}